\documentstyle[12pt]{article}
\begin{document}
\noindent
Jacques Demaret \par \noindent
Universit\'e de Li\`ege \par \noindent
Institut d' Astrophysique \par \noindent
Av. de Cointe 5, B-4000 Li\`ege \par \noindent
Belgique \par
\vspace{0.3cm}
\noindent
Michael Heller \par \noindent
Vatican Observatory \par \noindent
V-00120 Vatican City State \par
\vspace{.3cm}
\noindent
and \par
\vspace{.3cm}
\noindent
Dominique Lambert \par \noindent
Facult\'es Universitaires N. D. de la Paix \par \noindent
Facult\'e des Sciences \par \noindent
61 rue de Bruxelles, B-5000 Namur \par \noindent
Belgique \par
\vspace{.7cm}
\begin{center}
{\Large \bf LOCAL AND GLOBAL PROPERTIES OF THE WORLD}
\end{center}
\par
\vspace{0.5cm}
{\bf Key Words:} Noncommutative geometry, Quantum gravity,
Ultimate physical theory.
\vspace{0.5cm}
\begin{quotation}
{\em Abstract.} The essence of the method of physics is
inseparably connected with the problem of interplay between
local and global properties of the universe. In the present
paper we discuss this interplay as it is present in three major
departments of contemporary physics: general relativity, quantum
mechanics and some attempts at quantizing gravity (especially
geometrodynamics and its recent successors in the form of
various pregeometry conceptions). It turns out that all big
interpretative issues involved in this problem point towards the
necessity of changing from the standard space-time geometry to
some radically new, most probably non-local, generalization.
We argue that the recent noncommutative geometry offers
attractive possibilities, and give us a conceptual insight
into its algebraic foundations. Noncommutative spaces are, in
general, non-local, and their applications to
physics, known at present, seem very promising. One would expect
that beneath the Planck threshold there reigns a
``noncommutative pregeometry'', and only when crossing this
threshold the usual space-time geometry emerges.
\par
\end{quotation}
\vspace{0.5cm}

\section{Introduction}
Editors of a book devoted to the Mach Principle write in the Introduction:
\begin{quotation}
It is often not sufficiently appreciated how kind nature has
been in supplying us with `subsystems' of the universe which
possess characteristic properties (literally in the sense
`proper to the system') that can be described and measured
almost without recourse to the rest of the universe. (Barbour
and Pfister 1995)
\end{quotation}
Physics started its triumphant progress when people like Galileo
and Newton succeeded in isolating free fall of a stone from the
network of interactions shaping the structure of the
world. On the other hand, the question imposes itself: Is the
``whole of the universe'' a sum of its parts (or aspects) or
perhaps ``something more'', something that cannot be
reconstructed by investigating only ``local details''? It seems
that the essence of the method of physics is inseparably
connected with the problem of interplay between local and global
aspects of the world's structure. The aim of the present paper
is to discuss this interplay as it reveals itself in three major
departments of contemporary physics: general relativity, quantum
mechanics and some attempts at quantizing gravity.
\par
The very notion of local property is strictly connected with the
concept of point and its neighbourhood. No wonder, therefore,
that our analyses will focus on space-time structures. General
relativity is {\em par excellence\/} a theory of space-time.
Although its field equations, being differential equations, are
defined locally, their local character is of a very peculiar nature:
since Einstein's equations themselves determine the structure of
space-time on which they act, they are
intimately connected with the topological structure of the
underlying manifold which in turn should be taken into account
when solving the boundary condition problem. All these questions
have clearly a
global significance. We deal with them in Section 2.
\par
Non-relativistic character of quantum mechanics manifests itself
(among others) in a strong asymmetry of space and time (in
quantum mechanics there is
a position operator but there is no time operator). Some
attempts to cure this situation are physically interesting but,
at least for the time being, mathematically non-satisfactory.
The picture becomes even more complicated if one takes into
consideration the fact that two quantum systems (particles)
somehow know of each other, independently of the distance
separating them, as long as they were ``correlated'' in the
past. This ``non-separability'' effect strongly suggests that at
the more fundamental level the ordinary space-time geometry
breaks down and some new aspects of the local-global
interaction should be expected. This set of problems is
discussed in Section 3.=20
\par
One of the most ambitious attempts to reconstruct the space-time
geometry (or its substitute) at the fundamental level was known
under the name of {\em geometrodynamics\/} (Wheeler 1968, DeWitt
1967). The idea consisted in combining together the
spatio-temporal description of general relativity with the
probabilistic formalism of quantum mechanics. The set of all
3-geometries (called {\em superspace\/}) forms the arena of this
fluctuating geometry (or a quantum foam), and the probability
for a given 3-geometry to be the actual state of the universe
should be computed from the  so-called {\em Wheeler-DeWitt
equation\/}. When this idea had met serious difficulties Wheeler
(1980) proposed a new program to recover the macroscopic
space-time from what he called {\em pregeometry\/}, a stuff of
physics at the fundamental level.  Various entities (shapeless
collection of points, calculus of propositions, elementary acts
of measurements) were proposed as candidates for pregeometric
elements.  These rather vague ideas gave the beginning to a
number of mathematically sophisticated models. The situation in
this field is critically reviewed in Section 4.
\par
The above signalled attempts at penetrating the fundamental
level of physics suggest that at this level the concepts of
points and time instants loose their usual meaning and should be
replaced by some other mathematical structure. One such
mathematical structure has recently received the growing
interest, namely the so-called {\em noncommutative geometry\/}.
It is a vast generalization of the standard differential
geometry allowing one to investigate spaces which so far were
regarded as strongly pathological (e. g. non-Hausdorff spaces).
This is possible owing to the astonishing parallelism between
geometric and algebraic methods (discovered already by
Descartes). It turns out that this parallelism can be extended
to noncommutative algebras. In spite of the fact that
noncommutative spaces are, in principle, purely global
constructs, in which the concepts of points and their
neighbourhoods loose their usual meaning, the authentic dynamics can
be done on them (in terms of derivations of certain algebras as
counterparts of the usual notion of vector fields). Although
applications of noncommutative geometry to physics are still at
their preliminary stage, the obtained results are very
promising, and one branch of noncommutative geometry, the theory
of quantum groups, is now in the focus of interest of many
theorists. There are reasons to believe that at the fundamental
level it is a non-local physics (based on noncommutative
geometry) that governs the universe, and only above the Planck
threshold the ordinary (commutative) space-time geometry
emerges. In Section 5 we analyse the possibility of doing
geometry without local concepts. Applications of such a geometry
to fundamental physics are also briefly reviewed.
\par
In Section 6 we comment on a philosophical significance of the
above analyses.
\par
\section{Local and Global Aspects of the World in
General Relativity}

\subsection{Mach's Principle and General Relativity}
As it is well known, Einstein, in his way towards the theory of
general relativity, was greatly influenced by the set of ideas he
read out of Mach's writings, and which were called by him Mach's
Principle. Roughly speaking, Mach's Principle asserts that physical
properties, such as motion, inertia, centrifugal forces, must be
fully determined by the global structure of the universe
(distribution of masses in space). The following passages from
Mach's {\em Science of Mechanics\/} are often quoted as expressing
this doctrine:
\begin{quotation}
The universe is not {\em twice\/} given, with an earth at rest and
an earth in motion; but only {\em once\/}, with {\em relative\/}
motions, alone determinable (...) The principles of mechanics can,
presumably, by so conceived, that even for relative rotations
centrifugal forces arise.
\par
Newton's experiment with the rotating vessel of water simply
informs us, that the relative rotation of the water with respect to
the sides of the vessel produces {\em no\/} noticeable centrifugal
forces, but that such forces {\em are\/} produced by its relative
motion with respect to the mass of the earth and other celestial
bodies. No one is competent to say how the experiment would turn
out if the sides of the vessel increased in thickness and mass till
they were ultimately several leagues thick (...)
\par
When, accordingly, we say that a body preserves unchanged its
direction and velocity {\em in space\/}, our assertion is nothing
more or less than an abbreviated reference to {\em the entire
universe\/} (...) (Mach 1960)
\par
\end{quotation}
There are heated discussions, lasting to the present day (see
Barbour and Pfister 1995), as to whether, or to what extent, Mach's
Principle has been incorporated onto general relativity. Since the
outcome of these discussions depends on what does one precisely
mean under the name of Mach's Principle (there are many its
formulations and some of them are rather fuzzy), we shall not try
to multiply the possible answers. Instead, we shall adopt another
strategy. There is no doubt that general relativity exhibits a
subtle interplay of local and global properties of the universe,
and that this interplay is encoded in the mathematical structure of
this physical theory. Our goal will be to analyse the
mathematical structure of general relativity in order to
disentangle from it information about the interaction of local and
global properties of the world.
\par
\subsection{The Structure of Field Equations}
Field equations of general relativity are the result of the
encounter of two powerful Einstein's ideas. The first idea was
nicely encapsulated by Hermann Weyl in his known saying: ``space
tells matter how to move, and matter tells space how to curve''.
This, if suitably understood, is obviously  a postulate concerning
the interplay (a kind of feed-back) of the global properties of the
world (structure of space or space-time, large scale distribution
of matter) and its local properties (local curvature, description
of motion with respect to a local reference frame).
The second idea is that of geometrization of gravity, i.e. of a
``mapping of all the properties of the gravitational force and its
influence upon physical processes onto the properties of a Riemann
space '' (Stephani 1982, p. 82). Of course, both ideas are not
quite distinct: matter can tell space how to curve only if some
physical processes have been ``mapped'' into the geometry of space.
\par
Einstein's field equations are
$$
R_{\mu \nu} - \frac{1}{2}Rg_{\mu \nu } + \Lambda g_{\mu \nu } =3D
\kappa T_{\mu \nu }
$$
where the left hand side is purely geometric (we keep the
cosmological term $\Lambda g_{\mu \nu } $ for generality reasons)
and the energy-momentum tensor $T_{\mu \nu }$ on the right hand
side of these equations describes all forms of energy which can
produce a gravitational field. The above field equations constitute
a non-linear system of ten partial differential equations for
determining ten components $g_{\mu \nu }$ of the metric tensor
which are interpreted as gravitational potentials.
\par
Einstein's equations have a few properties which are important
from our point of view.
\par
First of all, even if we correctly choose the initial conditions
these equations have no unique solution, since it is always
possible to perform arbitrary coordinate transformations which do
not influence physical meaning of a solution (technically, the ten
above equations are not independent since the so-called
(contracted) Bianchi identities must be satisfied).
\par
Moreover, the field equations are not defined on an {\em a
priori\/} given metric space. More precisely, the vanishing of
divergence of the left hand side of Einstein's equations enforces
the vanishing of divergence of the energy momentum tensor $T_{\mu
\nu }$ (this fact is physically interpreted as the local conservation law).
But in order to compute the divergence of $T_{\mu \nu }$ one must
know the metric components $g_{\mu \nu }$. This is not a vicious
circle as it could look at first sight, but a deep aspect of the
interplay of local and global properties of the world as they are
encoded in the structure of the field equations.
\par
The above property is strictly connected with the non-linearity of
Einstein's equations. Owing to it the combined
gravitational effect of two bodies is not equal to the sum of the
effects of each of these two bodies separately: the interaction of
these two bodies with each other and with the generated
gravitational field gives an essential contribution to the final
effect. To see what does happen, one often uses the method of
successive approximations: one assumes that the space-time geometry
is determined by a ``part'' of the energy-momentum tensor. The
``rest'' of it is called the test body; it is affected by the
gravitational filed, but it does not contribute to it. However, one
should remember that this is only an approximation. In fact, the
universe as modelled by Einstein's equations is a non-linear
holistic system. This is clearly seen in the problem of defining
gravitational energy in the framework of general relativity. It is
typically a non-local entity: ``gravitational potential energy
contributes (negatively) non-locally to the total energy, and
gravitational waves can carry (positive) non-local energy away from a
system'' (Hawking and Penrose 1996, p. 72).
\par
In the next Subsection we shall discuss some local and
non-local properties as they appear in general relativity.
\par
\subsection{Local and Global Problems in General Relativity}
Another crucial property of the Einstein field equations is that
they are {\em hyperbolic\/} partial differential equations (see,
for instance, Choquet-Bruhat 1968). This property is closely
related to the fact that the metric $g$, which is to be determined
by Einstein's equations, is a Lorentz metric (with the signature $-
,+,+,+$) rather than the more usual Riemann metric (with the
signature $+,+,+,+$). Owing to this property space-time of general
relativity (i.e. space-time the metric tensor of which satisfies
Einstein's field equations) is locally the Minkowski (or
pseudoriemannian) space-time rather than the more standard Euclidean
one. This is a strong constraint on the local structure of
space-time coming from the very nature of (pseudo)Riemannian space: the
tangent space at every point of the (pseudo)Riemannian space must
be flat (pseudo)Euclidean independently of the global topological
or metric structure of a given space.
\par
This simple geometric property has important consequences for the
physical interpretation. It is a geometric counterpart of
Einstein's Equivalence Principle: the fact that space-time is locally
always flat
(up to any desired precision) means that locally the
gravitational field can always be transformed away, and
consequently that the special theory of relativity is locally always
valid.
\par
However, the interaction of any locality with the global structure
of space-time is not trivial. Only in the case, when the curvature
of space-time vanishes, localities simply ``add together'' to form
the Minkowski space-time (but even in this case one can change the
global topology by gluing together or cutting off certain parts of
space-time).
\par
In the next simple cases of space-times with constant curvature, or
with space-sections of constant curvature, interesting phenomena
can arise, such as the existence of closed timelike curves (in
space-times with constant curvature) or light cones starting to
reconverge (in space-times with space-sections of constant
curvature). The study of the global (or large scale)
structure of space-time has led to the formulation of many
problems, for instance:
\par
{\bf The problem of the chronological and causal structures of
space-time.} Two events $p$ and $q$ in space-time are said to be
{\em chronologically\/} or {\em causally\/} related if they can be joined by=
 an
oriented (piece-wise) smooth timelike or non-spacelike curve from
$p$ to $q$, respectively. Roughly speaking, a net of all such
curves joining all possible events in space-time forms
{\em chronological\/} and {\em causal structures\/} of space-time. The study=
 of
these two structures, interesting in itself, is an efficient tool
in investigating topological and other global properties  of
space-times (Carter 1971; Hawking and Ellis 1973, chapter 6; Beem and
Ehrlich 1981, chapter 2; Joshi 1993, chapter 4). It is interesting to
notice that the causal structure of space-time is closely connected with
the existence of a global time in the universe (i. e. time which
would measure the entire history of the universe). As it is well known,
there exist space-times which cannot be covered by a single coordinate
system (e. g. space-times with the topology of sphere), and consequently
no global time can be defined in such space-times. The necessary and
sufficient condition for the existence of a global time is the causal
stability of a given space-time. Space-time is said to be {\em causally
stable\/} if a small perturbation of its Lorentz metric does not produce
in it the appearance of the closed timelike curves (see Hawking and Ellis
1973, pp. 198-201).
\par
{\bf The problem of Cauchy horizons and Cauchy developments.} Owing
to the hyperbolic character of Einstein's equations the Cauchy data
given at the initial hypersurface in space-time, in general, do not
propagate throughout the entire space-time, but the region of their
influence (the so-called {\em Cauchy development\/}) is limited
by {\em Cauchy horizons\/}. Their existence is clearly connected with the
possibility (or impossibility) to determine the solution to Einstein's=
 equations from
the Cauchy data (the initial value problem) and with the
deterministic properties of a given space-time (see, Hawking and
Ellis 1973, chapter 7; Fischer and Marsden 1979).
\par
{\bf The cosmological horizon problem}. The existence of null-cones
in the tangent spaces at each event in space-time, interpreted as
the existence of the limiting velocity of the propagation of=20
physical signals, implies that various observers can influence (can
observe), or be influenced by, in principle, limited subsets of
events in space-time. Boundaries of these subsets are called
{\em (cosmological) horizons\/} (one distinguishes {\em particle=
 horizons\/},
and {\em past\/} and {\em future event horizons\/}, see Rindler 1977,
Tipler et al.
1980). The existence of horizons imposes severe constraints on the
observational testing of cosmological models and creates the
consistency problems for the standard cosmology (see, for instance,
Kolb and Turner 1990, pp. 261-269; Roos 1994; Partridge 1995).
\par
{\bf The singularity problem\/}, perhaps the most difficult and
most fundamental problem of all other problems. Roughly speaking,
singularities are a boundaries of space-time at which the manifold
structure of space-time breaks down. The Big Bang singularity in
the Friedman-Lema=8Ctre world models and the central singularity in
the Schwarzschild solution are the most notable examples of
singularities. More technically, singularities are defined in terms
of incomplete non-spacelike (causal) geodesics: a space-time is
{\em singular\/} if there is in it at least one non-spacelike incomplete
geodesics. By using this definition (or rather a criterion of the
existence of singularities), Penrose, Hawking and others have
proven several theorems about the existence of singularities in a
broad class of space-times satisfying rather tolerant conditions
(see, Hawking and Ellis 1973, Beem and Ehrlich 1981, Tipler et al.
1980, Clarke 1993, Earman 1955).
\par
Let us take a closer look at this problem since in it many aspects
converge of the local and global structures of space-time. The root
of the difficulty is connected with the fact that the Lorentz
metric carried by space-time is not a metric in the topological
sense. One can define a topology in terms of the chronological
structure of space-time, the so-called Alexandrov topology, but
without additional stronger assumptions it does not coincide with
the manifold
topology (in order to change it into the manifold topology
the so-called {\em strong causality condition\/} must be assumed
which asserts
that no neighbourhood of any of points of a given space-time is
intersected by a non-spacelike curve more than once, see Hawking and
Ellis 1973, pp. 192-198, Lerner 1972). The natural idea would be to
define a singular boundary of space-time as its Cauchy boundary
(defined, as usual, in terms of Cauchy sequences), but this cannot
be done because space-time does not carry the uniform structure
which is necessary for doing so (see Gruszczak and Heller 1993). It
was an ingenious idea of Schmidt (1971) to define the Cauchy
boundary of the total space of the frame fibre bundle over
space-time (which carries the suitable uniform structure), and by
``projecting it down'' to space-time to construct the singular
boundary of it, the so-called {\em b-boundary\/} of space-time. This
construction was regarded as an elegant and physically adequate
definition of singularities. It came as a surprise when Bosshard
(1976) and Johnson (1977) demonstrated that in the closed Friedman
world model the initial and final singularities form the single
point of the b-boundary, and that in the closed Friedman and
Schwarzschild solutions their b-boundaries are not Hausdorff
separated from the rest. Later on, strong indications were provided
(Geroch et al. 1982) that this situation is fairly typical for a wider
class of singular boundary constructions. These difficulties have led to
``a tension between the noun and adjective'' understanding of
singularities. ``The former attempts to conceive of singularities as
entities that can be localized while the latter eschews localization and
is content to speak of singular spacetimes when these spacetimes exhibit
large-scale or global features'' (Earman, p. 28).
\par
The story has its continuation, some
aspects of which will be touched upon in the next Subsection, but
even now it is clear that in the situation when the standard
structure of space-time breaks down it is the interaction between
``local'' and ``global'' that is severely perturbed (the beginning
and the end of the world become the single point of the b-boundary,
space-time loses its usual Hausdorff separability properties). Is
this a pure pathology or perhaps an indication of some deeper
regularities?
\par
\subsection{Global Formulations of General Relativity}
A strategy to solve at least some problems connected with the interaction
between geometric structure of space-time and the large structure
distribution of matter could consist in entirely eliminating the concept
of space-time from the foundations of general relativity and deriving it
on later stages of its construction. Such a possibility was suggested
by Geroch (1972). Usually, the smooth manifold structure on a
(non-empty) set $M$ is defined in terms of a smooth atlas on $M$.
However, it is well known that it can equivalently be defined in
terms of the algebra $C^{\infty }(M)$ of smooth functions on $M$.
Moreover, the algebra $C^{\infty }(M)$ can be regarded as a primary
structure and the manifold $M$ as a derived structure, namely as the set
of characters of the algebra $C^{\infty }(M)$.\footnote{Elements of the
(algebraic) dual space $C^{\infty }(M)^*$ with respect to $C^{\infty }(M)$
are called {\em characters\/} of the algebra $C^{\infty }(M)$.} When this=
 strategy is
applied to general relativity, the smooth functions, elements of the
algebra $C^{\infty }(M)$, can be identified with scalar fields.
Moreover, as shown by Geroch (1072), the Einstein field equations can be
written as functional equations in terms of $C^{\infty }(M)$. In this
way, we have a global formulation of general relativity in which scalar
fields are a primary concept and space-time a derived one.
\par
Since, however, Geroch's formulation of general relativity is
equivalent to its standard formulation we obtain nothing really new,
except for the fact that Geroch's formulation is open for further
generalizations.
\par
The next logical step would be to take any functional algebra $C$, and
to treat functions belonging to $C$ {\em ex definitione\/} as smooth
functions on the space of characters of $C$. Such a space (satisfying
two additional conditions, namely the closeness with respect to
localization and closeness with respect to composition with the
Euclidean functions) is called a {\em differential space\/} (for details see
Gruszczak et al. 1988). Owing to the two above mentioned conditions
differential geometry can be done on differential spaces. In particular,
all quantities required to define Einstein's equations (curvature, Ricci
tensor...) can be defines in terms of $C$. Differential spaces
satisfying Einstein's equations are called {\em Einstein algebras\/}
(see Heller 1992). Since the concept of smoothness is here generalized
as compared with the standard differential geometry, Einstein algebras
are authentically more general than the usual theory of general
relativity. In contrast with general relativity some weaker types of
singularities can be fully described in terms of Einstein algebras.
\par
The further generalization consists in replacing the functional algebra
$C$ by a sheaf ${\cal C}$ of functional algebras; then differential space
is replaced by what is called {\em structured space\/} (see Heller and
Sasin 1995a); the corresponding sheaf ${\cal C}$ of functional algebras
is called a {\em differential structure\/} on a given structured space. By
defining Einstein algebras in terms of this differential structure one
obtains the {\em sheaf of Einstein algebras\/}. It turns out that all
sorts of singularities can be described in terms of structured spaces.
Even if in some stronger types of singularities the structured space
structure behaves badly, by using this approach one can fully analyse
the situation.
\par
As an example, let us consider the initial and final
singularities (understood as the b-boundary points) in the closed
Friedman world model. Let ${\bar M} =3D M\cup \partial_b M$ be the
b-completed space-time of the closed Friedman world model, where
$M$ is the space-time of this model, and $\partial_b M$ its b-boundary.
$M$ is open and dense in ${\bar M}$. Since $M$ is a smooth manifold, one
can easily describe it as a structured space with the corresponding
differential structure ${\cal C}$. It turns out that the differential
structure ${\cal C}$ can be extended to a differential structure
${\bar{\cal C}}$ on the b-completed space-time ${\bar M}$, but only in a
trivial way, i. e. only constant functions smoothly (in the generalized
sense) extend to ${\bar M}$. In the theory of structured spaces
derivations of ${\cal C}$ play the role of vector fields. Of course,
a derivation of a constant function vanishes. This means that only zero
vector fields extend to ${\bar M}$, and consequently that the ``bundle
length'' of all curves joining the initial and final singularities is
equal to zero. The b-completed space-time ${\bar M}$ of the Friedman
closed model shrinks to the single point (the Hausdorff separability
breaks down). The global structure of the Friedman universe behaves in a
strongly pathological way. However,  locally (i. e. if one restricts
oneself to $M$ or to its open subsets; we remember that $M$ is open in
${\bar M}$) everything is all right. (For the detailed analysis of this
situation see Heller and Sasin 1955a, b; in these works such situations
are called {\em malicious singularities\/}.)
\par
The above analysis clearly shows that it is an interaction between local
and global properties of space-time that is the main factor of the
singular behaviour notoriously met in general relativity.
\par

\section{Quantum Mechanics: Towards New Conceptions
of Time and Space?}

\subsection{Introductory Remarks}

In this Section we analyse fundamental concepts of quantum
mechanics. We show that they lead to some problems which force
us to modify the usual notion of space-time. The first problem
comes from the status of time in classical quantum mechanics.
There exists in fact a deep conceptual asymmetry between space
and time in quantum mechanics: space is quantized whereas time
is not. Thus, time is ``infinitely divisible''.  This leads to
``strange" consequences, for example to the so-called ``Zeno's
paradox" (see below, Section 3.1). Its interpretation is
difficult because it is deeply related to the process of
measurement which is not completely understood in the framework
of the standard interpretation of Bohr's school.  This is partly
due to the irreversibility of the measurement process which
contradicts the reversibility of the evolution equation
(Schr\"odinger's equation, for example). To understand the
irreversibility of time we should introduce, as in Prigogine's
work, a time operator. However, it seems that, although this
work is intuitively very interesting, it is not entirely
satisfactory from the mathematical point of view. Nevertheless,
as we shall see, problems induced by the status of time in
quantum mechanics suggest a modification of its mathematical
nature. The second problem is related to the famous E.P.R.
paradox which introduces the idea of non-locality or more
precisely of non-separability with respect to space. In fact, in
quantum mechanics space cannot be viewed as a set of isolated
points. These problems lead to a deep modification of our
representation of ``quantum" space-time. It is interesting to
notice that the above mentioned problems concerning the nature
of time and the problem of the pointlike structure of space-time
were in fact present in the debate between Einstein and Cartan.
\par

\subsection{Time and Quantum Mechanics}=20

In quantum mechanics, every system is described by a wave
function $\Psi$ which obeys Schr\"odinger's evolution equation
\begin{eqnarray*}
 H \Psi =3D \frac{i h}{2 \pi} \frac{\partial}{\partial \, t} \Psi
\end{eqnarray*}
where $H$ is the Hamiltonian and $h$ the Planck constant. The
Hamiltonian involves two terms: one is related to the kinetic
energy and the second describes the potentials associated with
the interactions. Thus, the Hamiltonian contains all dynamical
information concerning the evolution of the system.
\par
The Schr\"odinger equation allows us to connect the initial
value of the wave function $\Psi (0)$ with its value at an
arbitrary time instant $t$. This can be expressed by using the
evolution operator $U$
\begin{eqnarray*}
 \Psi (t) & =3D  & U (t) \Psi (0) \\
 U(t)     & =3D & \exp \left( - \frac{2 \pi i} {h} H t \right)
\end{eqnarray*}
The evolution of the wave function is thus completely
deterministic.  Furthermore, to each evolution described by the
Schr\"odinger equation, we can associate a reversed evolution
changing the sign of time but without changing the potential of
the external forces (see Fer 1977). It is then possible to show
that the probability density $\Psi^{*} \Psi$ goes back in time
and that the mean value of the momentum has the opposite sign.
This is what is called the ``microreversibility" of quantum
mechanics. It is the consequence of the mathematical
structure of the Schr\"odinger equation. If we choose a suitable
evolution equation (for instance a nonlinear dynamical equation),
the microreversibility could immediately disappear.
Louis de Broglie (1956, pp. 144-164) tried to introduce such
nonlinear equations to avoid conceptual difficulties of the
usual quantum theory but without real success.
\par
The time variable appearing in quantum mechanics is thus
completely reversible as it is in classical and in relativity
theories. Furthermore, in quantum mechanics, time is not
described by a Hermitian operator as it is the case as far as
usual observables (position, momentum,...) are concerned; in
other words, time is not quantized (time variable commutes with
all observables; there is no operator canonically conjugated to
time). Even in the relativistic quantum field theory, time
remains reversible. These observations lead to a great
difficulty related to the so-called ``measurement problem".
\par
When a quantum system is not observed, it is described by the
evolution operator $U(t)$ which is the unitary operator. But, when it
is subject to a measure operation, the state of the system is
obtained by the use of a ``projector" (which is not a unitary
operator) describing the ``collapse" of the wave function
(which is in general a linear superposition of states) to a
particular state. The collapse is not a phenomenon which could
be explained in the framework of the usual quantum mechanics. It
should be regarded as a ``trick" allowing one to obtain the state which
results from the measurement process.  But now comes the
problem. We have seen that Schr\"odinger's equation is time
reversible but the collapse of the wave  function is, from its very
nature, irreversible.  There is something strange in
quantum mechanics! The impossibility of this physical theory
to give a satisfactory interpretation of the measurement
irreversibility problem=20
is at the root of some conceptual problems.
We shall describe one of them called ``Zeno's paradox" (for a
very interesting discussion of this paradox, see Omn\`es 1994;
Zeno's paradox was introduced for the first time by Misra and
Sudarshan (1977)).
\par=20
The quantum version of Zeno's paradox is related to a
strange property of the \mbox{(quasi-)}continuously observed
systems. Let us explain the core of the argument proposed by
Misra and Sudarshan.
\par
We start with an unstable particle and we assume that it can be
continuously observed. One could immediately object that this
kind of observation is impossible both theoretically and
practically. In fact, this objection is easily ruled out because
on the one hand, as we have seen above, in the usual quantum
mechanics time is not quantized. This means that we can consider
time as a continuous variable. On the other hand, a
\mbox{(quasi-)}continuous observation of an unstable particle
can be actually performed --- although only in an approximative
sense --- by using detection techniques as tracks in bubble
chambers.  This is in fact a weak argument, but it does not
matter since after all we are considering the quantum Zeno
paradox  as a kind of {\em Gedankenexperiment\/}.
\par
If the usual quantum theory is complete, it must give the
probability of decay of the unstable particle considered above
{\it when it is continuously observed}.  Quantum mechanics tells
us that the probability of observing the decay of such an
unstable particle during the time interval $t$ is proportional
to $t^2$. Let us denote by $p(t)$ the probability that we do not
observe any decay.  Then $p(t)$ can be written as
\begin{eqnarray*}
 p(t) =3D 1 - q(t) =3D 1 - a t^2,
\end{eqnarray*}
where $a$ is a real constant. If we make $n$ identical
observations during the time interval $t$, we can express the
probability of finding no decay after time $t$ in the following way
\begin{eqnarray*}
 p(t, n) =3D {\left( 1 - a {\left( \frac{t}{n} \right)}^2 \right)}^n
\end{eqnarray*}
Here we have assumed that each observation during the time
interval $t/n$ is independent of all other such observations.
Now, let us consider a continuous observation. We have to take
the limit of $p(t,n)$ when $n$ tends to infinity. But here the
paradox appears, because:
\begin{eqnarray*}
 \lim_{n \rightarrow \infty} p (n, t) =3D \lim_{n \rightarrow \infty}
 \exp \left( - \frac{t^2}{n} \right) =3D 1
\end{eqnarray*}
This is surprising because it means that an unstable particle
which is continuously observed will never decay. It is frozen in
its initial state by the fact of observation.  The decay is
proved to be impossible in the same way as the motion of the
arrow was shown by Zeno to be impossible. Misra and Sudarshan
have demonstrated that this paradox has a nice consequence for
the so-called ``Schr\"odinger's cat experiment".  A cat is
placed in a sealed box. In this box there is a system containing
a lethal gas which can be diffused if an unstable atom decays. Usual
quantum mechanics says that if we do not observe the cat, its
wave function is a superposition of two states: ``the cat alive"
and ``the cat dead". When we open the box, the collapse of the
wave function occurs and this superposition disappears: we have
only one of the two states quoted above.  Now, it turns out that
this explanation is too simple because due to Zeno's paradox, if
the cat observes continuously the system, he can stay alive!
\par
We could suspect that the difficulty of understanding Zeno's
paradox comes from the fact that we have no completely coherent
explanation of what measurement is in quantum mechanics. In
particular, we do not adequately understand the irreversibility
implied by the measurement. The paradox shows that an
irreversible process, such as the decay of a particle, is not
possible in the framework of the continuous observation.  But
why is this so? There are in fact at least two possibilities: (1)
The prediction coming from Zeno's paradox is true, i. e. Zeno's
paradox is not really a paradox but a relevant theoretical
result.  But then we have to explain why observation forbids the
decay.  And this is really difficult since the act of
measurement is not effectively described in the standard quantum
mechanics. (2) There could be that (quasi-)continuously
observed unstable systems effectively decay.  In this case, the
paradox would show that quantum mechanics is not complete
because it cannot allow us to compute the decay probability.
\par
The experimental test performed by Cook (1988) has shown that
Zeno's paradox is a true physical effect. Therefore, the second
possibility has been ruled out. But in order to understand this
effect, we have to consider not only an isolated system (the
unstable particle) but also a system strongly coupled to its
environment (see Joos 1996). This kind of approach is treated in
the framework of  the so-called ``decoherence theories". These
theories try to explain the emergence of the classical world
from the quantum one by a process which destroys the quantum
coherence through a strong coupling of a system to its
environment.  Unfortunately, however, we are not sure that these
theories offer any explanation of the reduction of the
wave-packet, i.e. of the ``measurement problem".
\par
It is therefore reasonable to believe that Zeno's paradox, and
maybe other classical paradoxes (``Schr\"odinger's cat",
``Wigner's friend",...) as well, give us a warning that one has
to modify the mathematical nature of time in order to get the
satisfactory interpretation of irreversible processes in the
quantum context. These processes are to be considered at two
different levels: physical irreversible processes (e. g.
processes of decay), and measurement processes (collapse of the
wave function). These two levels are conceptually distinct in
the standard interpretation of quantum mechanics.  We should
notice that the understanding of what we call irreversible
phenomena in classical mechanics does not directly imply the
modification of the mathematical nature of time (this can be
easily seen by inspecting simple models such as the ``Kac ring
model", see Kac  1959, p. 99). But nevertheless, for the
understanding of measurement process in quantum mechanics a
modification of the mathematical nature of time (and perhaps
also of its philosophical nature) would be required.
\par
This modification could be implemented by introducing a time
operator belonging to a noncommutative algebra. Even in the
context of classical dynamical systems, Prigogine and its school
have introduced such an operator (see Prigogine 1980; Prigogine
and Stengers 1984; Prigogine 1995a); it does not commute with
the evolution operator. Prigogine's intuition is that the
modification of the mathematical nature of time could explain
the fundamental nature of the irreversible phenomena. The main
idea of his theory is that we have to eliminate the concept of
individual trajectory by using probabilistic distributions.
This procedure has a nice consequence: it introduces a kind of
non-locality in space-time. Prigogine and Elskens (1985) write
\begin{quotation}=20
Irreversibility leads to a well-defined form of non-locality in
which a point is replaced by an ensemble of points according to
a new space-time geometry determined by the inclusion of the
privileged arrow of time.
\end{quotation}
Arguments used by Prigogine and his school are not yet completely
satisfactory at the mathematical level
(for a thorough discussion of Prigogine's arguments
see Bricmont 1995, pp. 159-208; Prigogine's (1995b) answer to
this paper can be found in the same issue of {\em Physicalia
Magazine\/}, pp. 213-218), but his intuition is interesting: a
modification of the mathematical nature of time, introduced in
order to understand the irreversibility of some processes, could
lead to a non-local character of space-time. We could think that
at the quantum level the introduction of a time operator would
force us to consider space-time as a non-local entity, i.e. a
geometrical object which is not definable starting from the
concept of ``point".
\par
\subsection{Non-Separability of the Quantum World}
First of all, we must distinguish ``non-locality" and
``non-separability".  Following Omn\`es (1994, p. 399),
non-locality characterizes a connection between two physical
systems which arises instantaneously irrespective of any
distance. More generally, we could admit that this connection is
realized via a space-like vector (but not necessarily
instantaneously). The non-separability says nearly the same thing
but here  ``one insists upon the impossibility of considering a
particle independently of the other one, as long as they are
strongly correlated in view of a common event in the past"
(Omn\`es, 1994, p. 399).
\par
From the well-known discussions around the E.P.R. paradox and
the Bell inequalities (see Jammer 1974, pp. 302-312), we have
learned that quantum mechanics involves a kind of
non-separability. Two systems which have interacted in the past
are correlated in the following sense. If we perform a
measurement on one of these systems it immediately affects the
second system independently of the distance between them. In
other words, the collapse of the wave function is really
instantaneous. Of course, this is due to the fact that the two
systems are described by a same wave function which spreads over
the whole space.  The connection between two correlated systems
is not the usual one, namely it is not a new kind of physical
interaction which would transmit some information or energy.
This non-separability occurs in all versions of quantum theory.
For example, if we consider Bohm's theory (see Jammer, pp.
278-296) or Nelson's (1985) stochastic quantum theory,  we are
lead to non-local potential and non-local effects as well.
\par
How is it possible to conceive such a non-separability between
two systems having interacted in the past? If we want to save
the relativity principle, i.e.  the Lorentz covariance, and the
usual causality, it is not possible to describe the E.P.R.
correlations between the systems using the properties of
(Minkowskian or even Riemannian) space-time. Therefore, we have
two types of interactions between physical systems. One is
described by the propagation of a signal on space-time according
to the laws of relativity, and the second is an instantaneous
correlation,  whatever the distance separating the systems,
which affects only the systems having interacted in the past
(following terminology proposed by Omn\`es (1994), we say that
this type of correlation is selective). We would certainly feel
more comfortably if all interactions between the systems were
described by the same unified concept in the geometrical context
of space-time.
\par
To describe the instantaneous collapse of the wave function
without introducing non-local influences (non-local potential as
in Bohm's theory), which would destroy Lorentz covariance, we
could think about a deep modification of the geometrical
structure of space. Let us suppose that space is no longer based
on point-like entities. Then one could consistently imagine some
type of non-separability which would be perfectly well described
in geometrical terms. Of course, such a new theory should give
the standard theory of general relativity as some sort of
approximation.
\par
As we have seen, quantum mechanics persuasively suggests the
necessity of modifying the nature of time, and quantum
correlations even more strongly compel us to look for a
drastically new concept of space which would be able to render
understandable instantaneous effects of the irreversible
collapse of the wave function.  Since, however, quantum
correlations are not present between all physical systems but
only between those systems which have interacted in the past,
the new geometrical structure of space-time should unify local
and non-local properties.
\par

\subsection{Back to the Past}=20
It seems that the new geometrical framework, suggested by the
problems arising in the standard formulation of quantum
mechanics, should not be founded on the concept of point as its
basic ingredient. Moreover, within the new framework one must be
able to consistently describe the time irreversibility of the
wave function collapse. Both these requirements remind us
proposals which were put forward by Cartan in his work about
manifolds endowed with absolute parallelism. It is known that
Cartan and Einstein discussed some extensions of general
relativity based on manifolds without curvature but with a
non-vanishing torsion. It should be remembered that such
manifolds admit different kinds of global parallelizations (a
detailed discussion of the theories based on the absolute
parallelism can be found in the book by Tonnelat 1965, pp.
274-288). In a famous note, Elie Cartan asks how is it possible
to restrict the class of such manifolds in order to describe
real physical phenomena.  He considers the case in which the
fundamental equations of a given theory remain invariant only
with respect to right-handed rectangular coordinates systems but
not with respect to the left-handed ones. This implies a kind of
fundamental irreversibility of the physical laws
\begin{quotation}
``On peut alors imaginer un syst\`eme d'\'equations $E$ qui
garderaient leur forme pour tous les syst\`emes de r\'ef\'erence
rectangulaires directs, mais qui changeraient de forme pour les
rep\`eres inverses. Un tel syst\`eme correspondrait \`a un
Univers dans lequel l'ensemble des lois du champ
gravitationnel-\'electromagn\'etique jouirait d'une esp\`ece de
polarisation: si on consid\`ere, par exemple, un syst\`eme de
charges \'electriques et leur \'evolution dans un certain
intervalle de temps, cette \'evolution serait impossible si on
renversait le sens de la dur\'ee: la physique serait {\em
irr\'eversible\/}. La th\'eorie classique ne pr\'esente rien de
pareil; mais il n'est pas interdit de penser que
l'irr\'eversibilit\'e de la Physique \'echappe \`a notre
exp\'erience, \`a cause de la faiblesse des champs qui entrent
dans notre domaine imm\'ediat de connaissance (Cartan 1974, p.
127)."
\end{quotation}
Irreversibility is thus connected to the intrinsic structure of
the space-time manifold. Now, Cartan has noticed that in
Einstein's theory of absolute parallelism there exist situations
in which it is impossible to give any meaning to the concept of
isolated physical corpuscle. In these situations we are forced
to abandon the individuality of physical points and we are lead
to a form of non-separability. As Cartan says it very clearly
\begin{quotation}
"...cette th\'eorie sera oblig\'ee de nier l'individualit\'e
physique des diff\'erents points qui constituent le fluide
\'electrique ou mat\'eriel suppos\'e \`a l'\'etat continu. Le
point mat\'eriel \'etait abstraction math\'ematique dont nous
avions pris l'habitude et \`a laquelle nous avions fini par
attribuer une r\'ealit\'e physique. C'est encore une illusion
que nous devons abandonner si la th\'eorie unitaire du champ
arrive \`a s'\'etablir (p. 128)."
\end{quotation}
\par
The theories with nonvanishing torsion developed by Einstein and
Cartan do not seem today very satisfactory, but they show something
which is intuitively very interesting: the irreversibility and
the non-separability can be obtained in a purely geometrical setting.
It is thus not unreasonable to think that quantum correlations
and all paradoxes related to the irreversible wave function
collapse could not be understood without a deep change of
geometrical ideas which lay at the basis of special and
general relativity theories in their standard formulations. A
non-local geometry would probably be needed to=20
unify the ideas of quantum mechanics with the theory of
gravitational field.
\par
\section{Pregeometry}

\subsection{Quantum Geometrodynamics and Superspace}

In the framework of general relativity, space-time is treated as
a continuum, i.e. as a four-dimensional pseudo-Riemannian
manifold.  The Hamiltonian formulation of Einstein's
geometrodynamics emphasizes the role of what Wheeler (1968) has
called \it superspace\rm, the arena wherein the curved space
geometry unfolds. Superspace is an infinite-dimensional space
each point of which constitutes a Riemannian 3-space
representing the space geometry of a relativistic space-time
that is to say a space-time which is a solution of Einstein's
classical field equations. The time evolution of such a
space-time, e.g.  a cosmological model, appears in superspace as
a continuous curve (with the precise beginning and end if the
model is closed). As strongly and repeatedly emphasized by
Wheeler, ``the dynamic object is not space-time: it is {\em
space\/}'' (Misner et al. 1973, p. 1181).

However, when dealing with distances of the order of the Planck
length, L$_{Pl}$ =3D $(\hbar G/c^3)^{1/2}$ =3D 1.6 $\times$
10$^{-33}$ cm, and less --- and we know that this could happen
at the very last stages of gravitational collapse or in the
close neighborhood of the initial or final cosmological
singularities --- quantum fluctuations take place in the
geometry of space and become predominant: accordingly, classical
geometrodynamics is superseded by  \it quantum
geometrodynamics\rm, initiated by Wheeler (1968) and DeWitt
(1967). Due to  Heisenberg's uncertainty principle,  in the same
way as in the usual quantum mechanics it is impossible to know
the position and the velocity of a particle at the same time,
one cannot know, in the framework of quantum geometrodynamics,
the precise 3-geometry and its rate of change at the same time
instant.  Space-time as a purely classical concept loses its
meaning and simply does not exist in the quantum gravity regime.

In cosmology, this implies some fuzziness in the geometry of the
universe which can now be described as a \it quantum foam\rm \ ---
a collection of quantum fluctuations (at the Planck scale),
continuously created and annihilated.  In such conditions,
elementary particles  of our familiar world  (protons, neutrons,
electrons...) should be regarded as gigantic excited states
travelling through the quantum foam, a picture as if directly
borrowed from Clifford's (1879)  anticipative
\it Space Theory of Matter\rm.

The deterministic classical history of space evolving in time is
now deprived of any meaning; one has to use explicitly the
language of probabilities and to speak of the probability that
the universe has actually such or such 3-geometry.  This
probability (more precisely, the wave function of the universe)
obeys the fundamental \it Wheeler-DeWitt equation\rm , the
gravitational counterpart of the famous Schr\"odinger's equation
of the usual quantum mechanics. The Wheeler-DeWitt equation is a
functional equation notoriously difficult to solve: its exact
solutions are known only for very symmetric cosmological models
(such as spatially homogeneous and isotropic
Friedmann-Lema\^{\i}tre models or anisotropic models of Bianchi
type) for which this equation reduces to a partial differential
equation. Even in these simplified situations, the answers to
important questions, such as the existence of the cosmological
singularity, remain ambiguous. This is due to considerable
technical difficulties within the formalism itself (see Misner
1969; Gotay and Demaret 1983), to the ignorance of the correct
boundary conditions which should be imposed on the wave function
of the universe (see Hartle and Hawking 1983; Vilenkin 1988)
and, not the least, complex --- and still unsolved ---
interpretative problems of the quantum formalism when it is
applied to the universe (see, for instance, DeWitt and Graham
1973).

Within the formalism of quantum geometrodynamics global time
does not exist any more: notions like ``before'' and ``after''
loose any meaning, and the concepts of space-time and time
appear only as valid in the classical approximation.
Consequently, they are secondary ideas in the formulation of a
fundamental physical theory.

Moreover, at a submicroscopic scale (more precisely below
Planck's scale), due to the inescapability of quantum
fluctuations in the 3-geometry, this geometry itself is not
deterministic any more; as expressed by Wheeler, ``it
`resonates' between one configuration (3-geometry) and another
and another'' (Misner et al. 1973, p. 1193).  Only when one
performs observations at a much larger scale do these quantum
fluctuations fit into a single space-time manifold, ruled by
Einstein's classical field equations.
\par

\subsection{Towards Pregeometry}
Wheeler's dream of building all physics on a purely geometric
basis, more precisely on quantum geometrodynamics, collapsed
when it became clear to himself that there was no natural place
inside this geometric formalism for spin 1/2 and in particular
for neutrinos (see Wheeler 1962, 1968).  Indeed, elementary
processes, such as pair creation, require a change in the
topology of 3-geometries (the topology of the initial 3-geometry
should develop a new wormhole to accommodate the new spin
structures associated with the created particles). However, such
a change in the topology of 3-geometries is totally forbidden
within the formalism based on classical differential geometry,
whose axioms are incompatible with the required
multiple-connectedness of space at Planck's scale.

In this way, the idea that geometry should constitute ``the
magic building material of the universe'' had to collapse on
behalf of what Wheeler has called \it pregeometry \rm (see
Misner et al. 1973, pp. 1203-1212; Wheeler 1980), a somewhat
indefinite term which expresses ``a combination of hope and
need, of philosophy and physics and mathematics and logic''
(Misner et al. 1973, p.  1203).

This fundamental change of perspective about the role of
geometry in the description of the physical universe is not
without link with considerations put forward by Sakharov as
early as in 1967. His point of view was that geometry should be
to elementary particle physics what elasticity is to atomic
physics. As elasticity cannot explain atoms, but, on the
contrary, atoms explain elasticity, geometrodynamics is not able
to explain particles: a particle built out of geometry would
look as queer as an atom made of elasticity. At a deeper level,
there should exist something --- call it \it pregeometry~\rm ---
which should account for geometry and which should certainly be
as removed from geometry as the quantum mechanics of atomic and
molecular systems is from elasticity.

Being deprived of any reference to the fundamental geometric
notions which constitute the heart of the theoretical
description of everyday physical reality, i.e. to space and
time, pregeometry --- whatever its precise formulation could be
--- is essentially a \it non-local \rm concept.
\par

\subsection{Many Faces of Pregeometry}
There is up to now no definite theoretical formulation of the
idea of pregeometry, but only a large variety of tentative
models more or less mathematically sophisticated (for a
bibliographical review of the fundamental properties of these
pregeometry models, see Gibbs (1995)).

However, many of these models do not consider, contrary to
Wheeler's point of view, space-time as an approximation to a
deeper and more fundamental substratum of a quite different
nature: they keep the idea of a preexisting space-time but view
it as a \it lattice\rm , i.e. as a \it discrete \rm structure
with the minimum length of the order of Planck length.
Philosophical and theoretical motivations for subscribing to
such an idea of a discrete space-time are quite diverse.

It is surely the advent of quantum mechanics, and especially the
discovery of the uncertainty principle, that led some
physicists, as early as in the 1930's, to speculate that
space-time could be discrete at the fundamental level.
Heisenberg (1930) himself had considered a lattice geometry to
try to get rid of the self-energy difficulty which plagued at
that time the electron theory, but he soon rejected it.  Some
years later, Einstein (1936) expressed the following opinion:
\begin{quotation}
\noindent
...perhaps the success of the Heisenberg method points to a
purely algebraic method of description of nature, that is, to
the elimination of continuous functions from physics. Then,
however, we must give up, by principle, the space-time
continuum...
\end{quotation}

Technical difficulties within the process of renormalization,
developed to eliminate the ultraviolet divergences present in
quantum field theory, have also reinforced the belief, held by
many physicists, in a natural cutoff at a very small length
scale.

A way of introducing the minimum length into physics has been
proposed by Snyder (1947). The replacement of space and time
coordinates by noncommutative operators leads to a quantization
of space-time, in consequence of the discrete nature of the
spectrum of these operators. Unfortunately, this model, although
Lorentz invariant, breaks the translation invariance. Similar
methods have been proposed later but without great success,
because of the difficulty of discretizing the full Poincar\'e
group.

Other attempts at building discrete space-time introduce
non-pointlike particles (superstring theory which views
particles as one-dimensional strings of Planck length might in
this respect be considered as the most advanced realization of
this program), or try to formulate the  field theory on a
discrete lattice (for references, see Gibbs (1995)).

Renewed interest in the possibility of quantizing space-time has
arisen in the framework of recent developments in the theory and
physical applications of \it quantum groups\rm , algebraic
structures which appear as deformations of the classical notion
of group (Gibbs 1995, pp. 25-28).

In all these tentative methods of developing pregeometry models,
one has to accept in advance a preexisting form of space-time.
This is not satisfactory from the point of view of Wheeler's
conception of pregeometry for whom the features of the
conventional space-time, such as its continuity, dimensionality,
and even causality and topology, should not be present from the
beginning but should naturally emerge in the transition process
from pregeometry to the usual space-time dynamics of our
conventional physical theories. The choice of appropriate
fundamental building blocks from which pregeometry is to be made
remains unspecified; this explains the large variety of
pregeometry models which have been proposed in the last decades.
Below we shall briefly describe  some of the most important and
original of these models.

\bf Wheeler's ``bucket of dust''. \rm In his first attempt to
formulate the concept of pregeometry, Wheeler (1964) discussed
the idea of ``dimensionality without dimensionality''. More
precisely, he asked whether geometry can be constructed out of
more basic elements, i.e.  out of a Borel set (a collection of
points (``bucket of dust'') devoid of any specific
dimensionality), when using the quantum principle. The hope
would be to  ascribe a probability amplitude to each possible
configuration of points in the Borel set and, in this way,
perhaps be able to explain why the dimensionality three would be
distinguished rather than any other dimensionality. But the
possibility of defining such a mathematical concept rests on
some notion of distance between two points, i.e. on a metric,
which is completely foreign to the idea of pregeometry.  As
noticed by Wheeler (1980, p. 3): ``Here also too much geometric
structure is presupposed to lead to a believable theory of
geometric structure.''

\bf Pregeometry as the calculus of propositions. \rm Afterwards
Wheeler (see Misner et al. 1973, pp. 1208-1212)  explored the
idea of using propositional logic (the logic of \it and, or \rm
and \it not \rm statements) as the fundamental building block of
pregeometry, space and time --- i.e. the continuum of everyday
physics --- hopefully emerging from the statistics of large
numbers of complex logical propositions. Why logic? Because, as
stated by Wheeler (Misner et al. 1973, p. 1212): ``Logic is the
only branch of mathematics that can `think about itself'.''

However, as shown somewhat later, this idea was not very
fruitful: mathematical logic does not appear as the natural
foundation for pregeometry: in order to give an account of
space-time, it is difficult to imagine how one could do without
any reference to the central principle of all physics, namely to
the \it quantum principle \rm (Patton and Wheeler 1975).

\bf Wheeler's self-reference cosmogony. \rm  Wheeler's latest
conception of pregeometry is deeply connected with the existence
of \it observers\rm.  Since the advent of quantum mechanics the
central role of the act of observation has been recognized: the
only way to say that an object exists or that a process is
taking place is to \it observe \rm it (``No elementary
phenomenon is a phenomenon until it is an observed (registered)
phenomenon'' (Wheeler 1979)). But the ultimate nature of any
measurement is a yes/no question posed by an observer: in the
case of the click of a counter, the information one deals with
in one yes/no bit of information (\it bit \rm or binary
digit is the basic unit of information), while in other types of
measurement large numbers of bits can be gathered (think, for
instance, of the registration of an interference pattern on a
screen). According to Wheeler, the universe is information
theoretical in nature, i.e. defined via discrete bits of
information. He expresses this idea in the following way: ``...
every physical quantity, every it, derives its ultimate
significance from bits, a conclusion which we epitomize in the
phrase It from Bit'' (Wheeler 1990).

But, in every measurement process, the observer acts on the
system he is studying and, in this way, he must play a role in
its future evolution.  The pure observer has been converted into
a participator: ``...in the elementary quantum phenomenon the
observer-participator converts conceivability into actuality''
(Wheeler 1980, p. 5).  Accordingly, the universe is \it
participatory \rm in its nature and the human observer is
endowed with an active and capital role of a participant in the
genesis of the universe (``Is observership the `electricity'
that powers genesis?'' (Wheeler 1977, p. 21)). Such a model is
called by Wheeler \it self-excited \rm and the corresponding
cosmogony is known as the \it self-reference cosmogony\rm : the
universe gives birth to communicating participators and
communicating participators give \it meaning \rm to the universe
through their continuous exchange of information (Patton and
Wheeler 1975, p. 565). This conception of the world is obviously
very near to the one advocated by the French philosopher Maurice
Blondel (1927): ``La pens\'ee cr\'e\'ee n'existe pas sans la
nature, et la nature elle-m\^eme se suspend \`a la pens\'ee
comme \`a sa raison d'\^etre\rm''. These ideas are also
manifestly deeply related to the Strong Anthropic Principle
(Barrow and Tipler 1985; Demaret and Lambert 1994) and to the
line of thought of the philosophical \it idealistic
\rm school with its famous representatives: Parmenides of Elea,
George Berkeley and the French philosopher Octave Hamelin who
tried to prove that the internal laws of the human cerebral
activity had \it necessarily \rm to give birth to the ensemble
of spatial, temporal, causal,... relations which  constitute
what we call the ``external world'' (see, for instance,
Gr\'egoire 1969, p. 54).

However, Wheeler's most recent conception of pregeometry should
not be too easily identified with the idealistic thought which
assigns the proper existence only to the mind. Wheeler's world
probably possesses some consistency on its own right; only
very few physicists would deny some reality to the world.

Up to now, nobody has succeeded in constructing a full
realization of Wheeler's proposal of pregeometry, i.e. an
information theoretical world defined by the participatory
observer. Indeed, such a task seems to be beyond our present
possibilities.

Other proposals for pregeometry emphasize the relational nature
of space-time.  The basic assumption common to all of them is
the hypothesis that there exist fundamental objects which can be
of different types: n-units (Penrose), preparticles (Bunge and
Garc\'{\i}a-Maynez, Garc\'{\i}a Sucre), quantum processes (von
Weizs\"acker, Finkelstein),...  Space-time would then consist of
the network of relations among these fundamental objects. We
give below very brief comments on these relational theories of
space-time (we refer to  Lorente (1993, 1995) and Gibbs (1995)
for more details).

The starting point of Penrose's (1971) model is an ensemble of
elementary objects called \it n-units\rm , each characterized by
the well-defined total angular momentum $n \times$ ${\hbar/2}$.
The interaction of all these objects between themselves gives
rise to the \it spin network\rm. There is no need for an
underlying space-time to begin with; space-time comes out at the
end. Penrose's ideas have been later elaborated by Ponzano and
Regge (1968) as well as by Hasslacher and Perry (1981) who have
shown that the quantum theory of gravity in three dimensions can
be described by means of the evaluation of spin networks by
explicitly using diagrammatic methods.  Recently, LaFave (1993)
has proposed a way of extending Ponzano-Regge model to four
dimensions by reinterpreting this theory in the light of
Wheeler's latest pregeometry philosophy.

Bunge and Garc\'{\i}a-Maynez (1977) and Garc\'{\i}a Sucre (1985)
have chosen as primitive concepts ``things'' not located in
space (which can be called \it preparticles\rm) acting among
themselves, the result of these interactions being identified
with the temporal and spatial structure of the world.

Von Weizs\"acker (1986) has considered the set of relations
among binary alternatives, called \it urs \rm  (equivalent to
yes/no experiments), at the basis of all quantum processes as
well as of space-time. In the same line of thought, Finkelstein
(1969-1974), in his series of papers about ``space-time code",
has considered the world as a network of quantum processes,
which he calls \it monads\rm, giving rise --- through their
interactions --- to space-time.

These relational theories of space-time are somewhat reminiscent
of Leibniz's conception of space and time as expressed in his
\it Monadology\rm .  According to Leibniz, space is but a set of
all ``points" (\it monads\rm) and of relations between them.

Some other pregeometric type of models are characterized by
abstract algebraic elements, the classical features of the world
emerging from this abstract system. In the model studied by
Cahill and Klinger (1996), called \it Heraclitean Quantum System
\rm (Heraclitus of Ephesus argued that the world is in the state
of flux and that the common sense is mistaken in regarding that
the universe is made of stable things), the algebra is taken to
be a Grassmann algebra (such an algebra is well-known from
modelling the fermionic sector of the standard model of
elementary particle physics).

Another purely algebraic attempt at modelling pregeometry  which
has received a great deal of attention in the last years is
based on noncommutative geometry\rm. The key idea of this model
is that the topological structure of space-time can be
understood in terms of essentially non-local mathematical
concepts, i.e.  in terms of a noncommutative algebra which would
play analogous role to the algebra of smooth functions on the
usual manifold.  The next Section of this essay will be devoted
to the overview of the foundations of this attractive new field
of mathematics which seems to be very promising for the study of
the quantum gravity regime of the universe below the Planck
threshold.
\par
\section{Non-Local Geometry and Non-Lo\-cal Phy\-sics}
\subsection{Introductory Remarks}
As we have seen in the preceding sections, in contemporary
physics many signals appear suggesting that on the fundamental
level time and space in their usual form might not exist and,
consequently, that the ``beginning'' of the universe might be
aspatial and atemporal. However, the models known so far said
very little how physics with no space and no time could be like.
In particular, no mathematical structures were known which could
adequately be able  to model such situations in their full
generality. All models used so far by physicists in this domain
were either approximate or toy models, or were based on a
non-fully understood mathematics (or both). The state of the art
has significantly changed after Alain Connes (together with his
co-workers) has elaborated a bundle of mathematical results
known under the common name of {\em noncommutative geometry}.
Although its applications to physics are still at their
preliminary stage, at least we have a sound mathematical theory
which is able to deal with entirely non-local situations. The
aim of the present section is to give a conceptual insight into
mathematical foundations of noncommutative geometry. This is
important from the philosophical point of view since by
penetrating into foundations of this geometry we could
understand how physics  {\em is possible \/} with no points in
space and no instants in time. Actual physical models based on
noncommutative geometry are for us here of secondary interest;
they will be only briefly mentioned in subsection 5.4.
\subsection{The Concept of Point}
It is sometimes said that space is collection of points. This
saying is misleading since it suggests that the concept of point
is not analysable, and this is not true. In the traditional
geometry the concept of point can be introduced (at least) in
four different ways. Although all these ways are equivalent, it
is worthwhile to enumerate them all, since in noncommutative
geometry they can lead to different generalizations.
\par
{\bf A.} Let $M$ be a smooth manifold. Usually $M$ is defined in
terms of a smooth atlas on a certain set but, as it is well
known, the entire smooth manifold structure is encoded in the
algebra $C^{\infty}(M)$ of smooth (real) functions on $ M$, and
the manifold can be equivalently defined in terms of this
algebra. Let $x\in M$ and let $F_x$ be the set of all functions
$f\in C^{\infty}(M)$ which vanish at $x$.  The sets $F_x$, for
every $x$, are maximal ideals in the algebra $ C^{\infty}(M)$.
It can be demonstrated that the existence of points in $M$ is
equivalent to the existence of maximal ideals in
$C^{\infty}(M)$.
\par
{\bf B.} An $*$-homomorphism (an involutive homomorphism) $
\chi :A\rightarrow {\bf C}$ from an
algebra $A$ into the field of complex numbers {\bf C} is said to
be a  {\em character\/} on the algebra $A$ (here and in what
follows we consider only involutive associative algebras with
units). It can be shown that=20
there exists a one-to-one correspondence between characters on
the algebra $C^{\infty} (M)$ and the maximal ideals of this
algebra and, consequently, the points of $ M$ are also
determined by characters on $C^{\infty}(M)$.
\par
{\bf C.} A linear functional $f$ on a $*$-algebra (involutive
algebra) $ A$ is called {\em positive\/} if $f(aa^{*})\geq 0$
for every $a\in A$. If, moreover, $ f(1)=3D1$, $f$ is called a
{\em state\/} on the algebra $A$. States which cannot be
presented as convex combinations of other states are said to be
{\em pure states\/} on $A$. It turns out that also pure states
on the algebra $C^{\infty}(M)$ uniquely determine the points of
$ M$.
\par
{\bf D.} To every state on the algebra $A$ there corresponds a
probability measure.  This probability measure for the algebra
$C^{\infty}(M)$ is of the Dirac's delta type.  As it can be
easily seen, it uniquely determines the points in $ M$.
\par
Clearly, the algebra $C^{\infty}(M)$ is commutative (since
pointwise multiplication of functions belonging to
$C^{\infty}(M)$ is a commutative operation), and it is precisely
this property of $C^{\infty}(M)$ that is closely connected with the
above methods of defining points in the manifold $M$. Moreover,
if an abstract algebra $A$ has maximal ideals (or, equivalently,
characters, pure states or Dirac's probability measures), on the
strength of the Gel'fand-Neimark-Segal (GNS) theorem it is
isomorphic to the functional algebra on a space $ M$, and the
points of $M$ can be determined be either of methods (A) -- (D).
\par
Dealing with commutative algebra $C^{\infty}(M)$ rather than
directly with the set $M$ opens the way for generalization. It
is natural to ask whether a noncommutative algebra $A$ could
also be interpreted as containing geometric information on a
certain space.
\subsection{Pointless Spaces}
As shown by Connes and his co-workers, the answer to the last
question of the preceding subsection is positive, although the
mathematics which must be invested in order to decipher the
geometric information contained in a noncommutative algebra is
rather complex. In the present subsection we shall first see how
the concept of point can disappear in noncommutative spaces, and
then how differential geometry and physics can be done on such
pointless spaces.
\par In the case of a commutative algebra $A$, there exists
the GNS isomorphism (for continuous functions) $A$$\cong
C^0({\rm M}{\rm a}{\rm x}\, A)$, where  ${\rm M}{\rm a}{\rm
x}\,A$ is the set of maximal ideals of the algebra $A$, given by
$a\mapsto\hat {a}$, for every $ a\in A$, $\hat {a}$ being the
mapping which sends each $a \in A$ to the mapping $$
\hat{a}: {\rm Max} A \rightarrow {\bf C}
$$ defined by
\[\hat {a}(I)=3Da+I\in A/I,\]
where $I\in {\rm M}{\rm a}{\rm x}M$. In the general case, when $
A$ is a noncommutative algebra, maximal ideals must be replaced
by {\em primitive ideals}, i. e. by kernels of irreducible
representations of $A$ in a Hilbert space. A {\em
representation\/} of an algebra $A$ in a Hilbert space ${\cal
H}$ is a mapping $
\rho :A\rightarrow {\rm E}{\rm n}{\rm d}\,{\cal H}$$ $ of the algebra $
A$ into the set of linear transformations of the Hilbert space $
{\cal H}$ (such transformations are called {\em endomorphisms\/}
of ${\cal H}$ or operators acting on $ {\cal H}$) preserving
essential properties of the algebra (addition and multiplication
of elements of $A$, and their multiplication by scalars). A
representation $
\rho\rightarrow {\rm E}{\rm n}{\rm d}\,{\cal H}$
is said to be  {\em irreducible\/} if only invariant subspaces
of $ {\cal H}$ are \{0\} and ${\cal H}$ itself, where by an {\em
invariant subspace\/}  of ${\cal H}$ one understands a subspace
${\cal H}_0\subset {\cal H}$ such that, for any endomorphism $
\rho (a)\in {\rm E}{\rm n}{\rm d}\,{\cal H}$, $a\in A$, one has $
\rho (a){\cal H}_0\subset {\cal H_0}$.
And finally, the kernel of the representation $\rho\rightarrow
{\rm E}{\rm n}{\rm d}\,{\cal H}$, Ker$\rho$, is defined as ${\rm
K}{\rm e}{\rm r}\rho :=3D\{a\in A:\;\rho (a)=3D0\}$. Let us notice a
certain similarity between the concept of maximal ideals and
that of primitive ideals: in defining maximal ideals we require
vanishing of functions at certain points of a set; in defining
primitive ideals we require vanishing of representation mappings
on certain elements of the algebra.
\par
Let us denote the set of all primitive ideals of $A$ by ${\rm
P}{\rm r}{\rm i}{\rm m}\, A$. If $A$ is commutative then ${\rm
P}{\rm r}{\rm i}{\rm m}\,A=3D{\rm M}{\rm a}{\rm x}\, A$, and we go
back to the previous construction. If $A$ is noncommutative we
also have a mapping
\[\hat {a}:\;{\rm P}{\rm r}{\rm i}{\rm m}\,A\rightarrow
A/P,\] for $P\in {\rm P}{\rm r}{\rm i}{\rm m}\,A$, given by
\[\hat {a}(P)=3Da+P\in A/P,\]
but the quotient algebra $A/P$ can be very complicated, for
instance the dimension of $A/P$ can change as $P$ changes. In
the case when $ {\rm P}{\rm r}{\rm i}{\rm m}\,A$ is a Hausdorff
space, there is a counterpart of the GNS isomorphism (see
Dupr\'e 1978). To articulate it let us construct a disjoint
union of the quotient algebras
\[E:=3D\bigcup_P\{A/P:P\in {\rm Prim}\,A\}\]
with a suitable topology, and define the bundle $\Omega
:=3D\{E,{\rm P}{\rm r}{\rm i}{\rm m}\,A,\pi \}$ where $\pi
:E\rightarrow {\rm P}{\rm r}{\rm i}{\rm m}\,A$ is an obvious
projection (it is, in fact, a Banach bundle). It can be shown
that the set $\Gamma^0(\Omega )$ of (bounded) continuous
cross-sections of $
\Omega$
forms a $C^{*}$-algebra, and one obtains the isomorphism of $ A$
onto the $C^{*}$-algebra $\Gamma^0_c(\Omega )$ of compactly
supported cross-sections of $
\Omega$, $A\cong\mbox{$\Gamma^0_c(\Omega )$}$. For
non-Hausdorff spaces more difficulties arise and the
construction is not that obvious (see Dupr\'e 1978).
\par
As we can see, in the case of noncommutative spaces (even for
Hausdorff topologies), the idea of a family of continuous
functions vanishing at a given point (maximal ideal of the
algebra $A$) is replaced by the kernel of an irreducible
representation of the algebra $A$ in a Hilbert space (primitive
ideal). However, we must remember that in many applications the
elements of the algebra $A$ are  cross-sections of a Banach
bundle, and consequently they are global entities.
\par
In the noncommutative case, we have the correspondence between
representations of the algebra $A$ in a Hilbert space $ {\cal
H}$ and states on the algebra $A$, and between irreducible
representations of $ A$ in ${\cal H}$ and pure states on $A$,
but of course the existence of pure states is no longer
equivalent to the existence of points in the considered space
(with a certain degree of tolerance, pure states could be
regarded as generalizations of points).
\par
However, it should be noticed that in some rather special
cases, a noncommutative algebra $A$ can admit maximal ideals. In
such a case, if ${\cal I }$ is a (two-sided) maximal ideal of $A$,
then the quotient $A/{\cal I }$ is a simple algebra (i.e. it has
no non-trivial two-sided ideals). These maximal ideals can be
regarded as ``points'' of the noncommutative space determinded
by the algebra $A$, but such ``points'' can have rich ``internal
structure''. One says that they ``take their values in a simple
algebra''. This remains in contrast with the commutative case where,
for a maximal ideal ${\cal I }$ of $A$, one has $A/{\cal I }
\simeq {\bf C}$, and one says that the points of the
corresponding space ``take their values in ${\bf C}$''. The last
property is the algebraic counterpart of the fact that in the
commutative space points have no ``internal structure''. (See
Masson 1996, pp. 91-97.)
\par
As an example of a space with ``structured points'' let us
consider the algebra $A =3D C^{\infty }(V) \otimes M(n, {\bf C})$
of smooth functions on a differentiable manifold $V$ with values
in the matrices $M(n, {\bf C })$. All such functions vanishing
at $p \in V$ form a maximal ideal ${\cal I }$ of $A$ which can
be identified with a point in a noncommutative space. Points of
these space ``take their values in the simple algebra $A/{\cal
I}$''.
\par
The existence of spaces with ``structured points'' opens
new possibilities as far as applications to physics are
concerned (see, for instance, a noncommutative version of the
Kaluza-Klein theory, Madore 1995, pp. 180-187).
\par
It turns out that noncommutative spaces, as defined by general
noncommutative algebras, are quite manageable provided we have
at our disposal rather sophisticated mathematical tools. For
instance, as has been demonstrated by Connes (1995), the measure
theory on noncommutative spaces is replaced by the theory of
von Neumann algebras, and many features usually dealt with by
using topological methods are captured by the K-theory. There
are also several ways of introducing differential calculus on
noncommutative spaces (they are transparently discussed by
Dubois-Violette (1995)). Unfortunately, we cannot enter here
into these interesting topics. The problem which is now
important for us is how physics can be done in terms of geometry
in which there could be no points in space  and no instants in
time.
\par
The essential thing for physics is dynamics, and any dynamics is
thought to be a process evolving in time. The standard way to
mathematically model dynamical processes is in terms of vector
fields. Solutions of the corresponding system of differential
equations (called {\em dynamical system\/}) give integral curves
of these vector fields which in turn are interpreted as
histories of the process. The value of the vector field at a
given time instant of the history is a vector (tangent to this
history) describing the ``behaviour'' of the system at the given
time instant. It should be noticed that although the concept of
vector is a local concept, and as such it could have no
counterpart in noncommutative geometry, the concept of vector
field is a global concept and it survives the generalization to
noncommutative geometry. It turns out to be enough to have a
``generalized dynamics'' in this new conceptual framework.
\par
A counterpart of a vector field in  noncommutative geometry is
a {\em derivation\/} of the algebra $A$, i. e. a mapping $
V:A\rightarrow A$ satisfying the Leibniz rule. In fact, one can
do differential geometry in terms of such derivations.  In
particular, connection, curvature, Ricci tensor, and
consequently Einstein (dynamical) equations can be defined (see,
Sasin and Heller 1995). However, one must remember that all
these concepts are non-local. For instance, curvature should not
be imagined as a ``curved space'' but rather as a certain
abstract operation on derivations of a given algebra. In the
noncommutative framework, one can also do differential geometry
in terms of (generalized) abstract differential forms rather
than in terms of derivations (see Madore 1995). In the cases
when both methods (in terms of derivations and in terms of
forms) are applicable one must adapt one's choice to the actual
situation.
\par
Non-commutative geometries become especially effective tool in
dealing with various ``pathological'' or ``singular'' spaces
(for instance, Penrose's tillings, foliated spaces) if the
algebra in question is a $C^{*}$-algebra. Connes (1995, chapter
2) has elaborated a method which allows one to convert a broad
class of noncommutative algebras into $C^{*}$-algebras. The
method consists in constructing a bundle the cross-sections of
which form an algebra. The suitable completion of this algebra
changes it into a  $C^{*}$-algebra.  Algebras of observables in
the standard formulation of quantum mechanics are the prototype
of  $C^{*}$-algebras (in fact, every  $C^{*}$-algebra can be
represented as a subalgebra of the algebra of such observables,
i. e. as a subalgebra of the algebra of bounded operators on a
Hilbert space). Exactly, because of that the theory of
$C^{*}$-algebras has been well developed and could be regarded
as a link between traditional mathematics and its
noncommutative generalizations.
\par
\subsection{Some Applications to Physics}
As we have noticed at the beginning of this section,
applications of noncommutative geometry to physics are at their
preliminary stage of development, but even at this stage they
are more than encouraging. One of the most important of these
applications is the result obtained by Connes and Lott (1990)
consisting in geometrizing the standard model of physical
interactions. As is well known, quantum electrodynamics with the
Maxwell-Dirac lagrangian gives a very elegant and very efficient
description of electromagnetic interaction. The standard model
generalizes this description to other interactions with the
exception of gravity. This model works very well but is not
elegant from the mathematical point of view: Its lagrangian is a
juxtaposition (a sum) of five terms, each of them describing a
contribution coming from a different source. Connes and Lott
have demonstrated that one can obtain an elegant
Maxwell-Dirac-like lagrangian for the standard model provided
one assumes that the underlying space-time, at the length scale
of the order of $10^{-16}$ cm, has the structure of a
noncommutative space, namely the structure of $M^ 4\times F$
where $M^4$ is a 4-dimensional manifold and $F$ is a space
consisting of two points, $F=3D\{a,b\}$ (in the framework of
noncommutative geometry this space can be given a metric
structure).
\par
Since the number of applications of noncommutative geometry to
physics rapidly increases, let us enumerate only some of them.
For example, methods of noncommutative geometry have been
applied to gauge theories (Dubois-Violette et al. 1990,
Chamseddine et al. 1992), unification theories (Chamseddine et
al. 1993a, Chamseddine and Fr\"ohlich 1994a), supersymmetry
theories (Chamseddine 1994), Chern-Simons theory (Chamseddine
and Fr\"ohlich 1994b), and to the hamiltonian formalism (Kalau
1996, Hawkins 1996). One of the present authors together with
his co-worker (Heller and Sasin 1996a, b) has used
noncommutative methods to study the problem of classical
singularities in general relativity.
\par
The most obvious idea would be to speculate that geometry
beneath the Planck threshold (i. e. in the quantum gravity
regime) is noncommutative with no space and no time in the
usual sense, and only by going to larger scales one would
obtain, via a kind of symmetry breaking, the standard
commutative geometry of space-time.  Unfortunately, before
implementing this attractive idea into a working mathematical
model some conceptual difficulties must be overcome. For the
time being some work has been done to generalize general
relativity to the noncommutative framework (Chamseddine et al.
1993b), and to couple gravity to the standard model of
fundamental interactions (Chamseddine and Connes 1996a, b).
\par
\section{Concluding Remarks}
As we have noticed in the Introduction, the very existence of
physics is strictly connected with the possibility of isolating
simple ``local subsystems'' from the net of entanglements
constituting the structure of the universe. Enormous successes
of the empirical method, based on this property, have somehow
overshadowed the fact that the strategy of isolating ``local
subsystems'' can be but an approximation to the more adequate
approach in the study of the world. Although the ``universe as a
whole'' always was a subject-matter of interest for many
physicists and astronomers, it was commonly believed that its
structure could be disclosed by investigating local physics as a
``fair sample'' of the rest. Even the beginnings of relativistic
cosmology were strongly biased by this prejudice.  There were
modern mathematical methods, used in general relativity, quantum
mechanics and quantum field theories, that gradually enforced
the new perspective.
\par
The typical feature of the 20th century mathematics is changing
from local methods, characteristic of the older approach, to
global methods of treating mathematical objects. Topology and
functional analysis, which have become standard tools of doing
mathematics, are global from the very beginning, and when
employed in other branches of mathematical investigation they
immediately produce problems with a pronounced global component.
This is especially evident in the domain of differential
geometry in which traditional ``differential methods'' give
often misleading results unless certain global conditions are
guaranteed. This effect is so strong that the usual stage for
differentially geometric investigation is nowadays not a
differential manifold itself but rather some ``larger spaces''
considered as global structures constructed over a given
manifold, such as foliations, fibre bundles or even families of
bundles (for instance, K-theory).
\par
It seems that noncommutative geometry is a theory in which the
above ``globalization process'' is at its apex. In a certain
sense, localities have been engulfed by the global structure of
noncommutative spaces, and they can only be recovered by
restricting the corresponding noncommutative algebras to some of
their subalgebras (for example to their centers).
\par
It goes without saying that such methods had sooner or later to
find their place in theoretical physics. The theory of general
relativity was perhaps the first physical theory upon which the
global approach has been enforced, but soon other physical
theories surrounded themselves to this new strategy. We have
observed the results of this process in the preceding sections.
\par
As we have also seen, there are reasons to believe that at the
fundamental level physics might be based on a noncommutative
mathematics. Preliminary results in this direction are
encouraging, and also ``philosophy'' of such an approach offers
attractive interpretative possibilities. Let us only mention two
widely discussed issues, which could find their unexpected
clarification within the framework of physics based on
noncommutative geometry, namely the Mach Principle problem (see
above section 2) and the non-separability of events in quantum
mechanics (section 3).
\par
All stronger formulations of Mach's Principle require that,
roughly speaking, local physics should be \it entirely
determined \rm by the global structure of the universe, and
general relativity (and other similar theories as well)
stubbornly fail to satisfy this requirement. Noncommutative
approach to physics at the fundamental level neatly clarifies
the situation. Beneath the Planck threshold there would be no
space-time in the usual sense but only a ``noncommutative
pregeometry'' with no non-trivial ``local neighbourhoods'' at
all (by ``trivial local neighbourhoods'' we mean those connected
with eventual commutative subalgebras of the corresponding
noncommutative algebra). In such circumstances, there would be
only the fully Machian physics entirely determined by the global
structure of the world. This ``Machian property'' should be
regarded as incorporated into the primordial symmetry, and the
present non-Machian physics as the result of the first symmetry
breaking in the history of the universe, i. e. of the transition
from noncommutative pregeometry to the usual commutative
space-time geometry.
\par
It is legitimate to assume that some ``fragments'' of the ``old
phase'' would remain frozen into the present structure of the
world. It seems reasonable to look for such vestiges of the
primordial non-local symmetry in the domain of microphysics
which is expected best to remember the broken primordial
symmetries. No wonder that they would somehow be encoded into
the structure of the phase space of quantum mechanics: all
information about two particles which once interacted with each
other is indeed contained in the same vector of the
corresponding Hilbert space. And this is irrespectively of how
great distance in space is separates them. After all, space
distance is the later concept which was not present in the
original symmetry.
\par
At the end a word of warning seems indispensable. Many
beautiful philosophies have collapsed because they were unable
to find their support in a solid physical theory. Whether the
looked for theory of ultimate physics will really be based on
noncommutative mathematics remains to be seen. The present
preliminary results, although far from being conclusive, do not
discourage such a belief.
\par
\vspace{1.8cm}
\noindent
\begin{center}
{\em References}
\end{center}
\par
\vspace{0.5cm}
Barbour, J. B. and Pfister, H. 1995, {\em Mach's Principle: From
Newton's Bucket to Quantum Gravity\/}, Birkh\"auser: Boston --
Basel -- Berlin.
\par
Barrow, J.D. and Tipler, F.J. 1986, \it The Cosmological Anthropic
 Principle\rm, Clarendon Press: Oxford.
 \par
Beem, J. K. and Ehrlich, P. E. 1981, {\em Global Lorentzian
Geometry\/}, M. Dekker: New York -- Basel.
\par
Blondel, M. 1927, \it La Pens\'ee\rm , cited by: C. Tresmontant 1963, \it
Introduction \`a la m\'etaphysique de Maurice Blondel\rm , p. 63, \'Editions
 du Seuil: Paris.
\par
Bosshard, B. 1976, {\em Commun. Math. Phys.\/}, {\bf 46}, 263.
\par
Bricmont, J. 1995, Science of Chaos and Chaos in Science, \it Physicalia
Magazine\rm , \bf 17\rm, 159.
\par
Bunge, M. and Garc\'{\i}a-Maynez, A. 1977, A Relational Theory of Physical
Space, \it Int. J. Theor. Phys., \bf 15\rm , 961-972.
\par
Cahill, R.T. and Klinger, C.M. 1996, Pregeometric Modelling of the
Space-Time Phenomenology, preprint gr-qc/9605018.
\par
Cartan, E. 1974, Le parall\'elisme absolu et la th\'eorie
unitaire du champ, in:
\it Notices sur les travaux scientifiques\rm, Gauthier-Villars, Paris.
\par
Carter, B. 1971, Causal structure in Space-Time, {\em Gen. Rel.
Gravit.\/} {\bf 1}, 349-391.
\par
Chamseddine, A. H. 1994, Connection Between Space-Time
Supersymmetry and Non-Commutative Geometry, {\em Phys. Lett.\/},
{\bf B332}, 349-357.
\par
Chamseddine, A. H. and Connes, A. 1996a, {\em The Spectral Action=
 Principle},
preprint.
\par
Chamseddine, A. H. and Connes, A. 1996b, {\em A Universal Action Formula},
preprint.
\par
Chamseddine, A. H., Felder, G. and Fr\"ohlich, J. 1992, Unified
Gauge Theories in Non-Commutative Geometry, {\em Phys. Lett.\/},
{\bf B296}, 109-116.
\par
Chamseddine, A. H., Felder, G. and Fr\"ohlich, J. 1993a, Grand
Unification in Non-Commutative Geometry, {\em Nuclear Phys.\/}
{\bf B395}, 672-698.
\par
Chamseddine, A. H., Felder, G. and Fr\"ohlich, J. 1993b, Gravity
in Non-Commutative Geometry, {\em Commun. Math. Phys.\/} {bf
155}, 205-217.
\par
Chamseddine, A. H. and Fr\"ohlich, J. 1994a, SO(10) Unification
in Noncommutative Geometry, {\em Phys. Rev.\/}, {\bf D50}, 2893-2907.
\par
Chamseddine, A. H. and Fr\"ohlich, J. 1994b, The Chern-Simons
Action in Noncommutative Geometry, {\em J. Math. Phys.\/}, {\bf
35}, 5195-5218.
\par
Choquet-Bruhat, Y. 1968, Hyperbolic Partial Differential Equations
on a Manifold, in: {\em Battelle Rencontres\/}, eds. C.M. DeWitt
and J.A. Wheeler, Benjamin: New York -- Amsterdam, pp. 84-106.
\par
Clarke, C. J. S. 1993, {\em The Analysis of Space-Time
Singularities\/}, Cambridge University Press: Cambridge.
\par
Clifford, W.K. 1879, \it Lectures and Essays\rm , vol. \bf 1\rm, eds.
L. Stephen and F. Pollock, Macmillan: London.
\par
Connes, A. 1995, {\em Noncommutative Geometry}, Academic Press: New York,
London, etc.
\par
Connes, A. and Lott, J. 1990, Particle Models and Noncommutative Geometry,
{\em Nuclear Phys.\/} {\bf 18B}, suppl. 29-47.
\par
Cook, R.J. 1988, What are Quantum Jumps?, \it Physica Scripta\rm, \bf 21\rm,
49.
\par
de Broglie, L. 1956, \it Nouvelles perspectives en microphysique\rm, Albin
 Michel, Paris.
\par
Demaret, J. and Lambert, D. 1994, \it Le principe anthropique\rm ,=
 \'Editions
Armand Colin: Paris.
\par
DeWitt, B.S. 1967, Quantum Theory of Gravity I. The Canonical Theory,
\it Phys. Rev.\rm, \bf 160\rm, 1113-1148.
\par
DeWitt, B. S. and Graham, N.G. (eds.) 1973, \it The Many-Worlds
 Interpretation
of Quantum Mechanics\rm , Princeton University Press: Princeton.
\par
Dubois-Violette, M. {\em Some Aspects of Noncommutative Differential=
 Geometry},
preprint, q-alg/9511027.
\par
Dubois-Violette, M., Kerner, R. and Madore, J., 1990, {\em J.
Math. Phys.\/}, {\bf 31}, 323-330.
\par
Dupr\'e, M. 1978, Duality for  $C^{*}$-algebra{\em s,\/} in: {\em=
 Mathematical
 Foundations of
Quantum Theory}, Academic Press: New York, London, etc.
\par
Earman, J. 1955, {\em Bangs, Crunches, Whimpers, and Shrieks:
Singularities and Acausalities in Relativistic Spacetimes\/}, Oxford
University Press, New York -- Oxford.
\par
Einstein, A. 1936, Physics and Reality, \it Journal of the Franklin
Institute\rm , \bf 221\rm , 378.
\par
Fer, F. 1977, \it L'irr\'eversibilit\'e, fondement de la stabilit\'e du
 monde
physique\rm, Gauthier-Villars, Paris.
\par
Finkelstein, D. 1969, Space-Time Code I, \it Phys. Rev., \bf 184\rm ,
 1261-1271
and the subsequent papers of the same series in \it Phys. Rev. D \rm
between 1972 and 1974; cf. also Finkelstein, D. 1996, \it Quantum
Relativity\rm , Springer-Verlag: Berlin.
\par
Fischer, A. E. and Marsden, J. E. 1979, The Initial Value Problem
in the Dynamical Formulation of General Relativity, in: {\em
General Relativity -- An Einstein Centenary Survey\/}, eds. S. W.
Hawking and W. Israel, Cambridge University Press: Cambridge --
London, etc.
\par
Garc\'{\i}a Sucre, M. 1985, Quantum Statistics in a Simple Model of=
 Space-Time,
\it Int. J. Theor. Phys.\rm , \bf 24\rm , 441-445.
\par
Geroch, R., 1972, {\em Commun. Math. Phys.\/}, {\bf 26}, 271.
\par
Geroch, R., Liang Can-bi and Wald, R. M. 1982, Singular Boundaries of
Space-Tiles, {\em J. Math. Phys.\/}, {\bf 23}, 432-435.
\par
Gibbs, P. 1995, The Small Scale Structure of Space-Time: A Bibliographical
Review, preprint hep-th/9506171.
\par
Gotay, M.J. and Demaret, J. 1983, Quantum Cosmological Singularities, \it
 Phys. Rev. D\rm , \bf 28\rm , 2402-2413.
\par
Gr\'egoire, G. 1969, \it Les grands probl\`emes m\'etaphysiques\rm , Presses
Universitaires de France: Paris.
\par
Gruszczak, J., M. Heller, and Multarzy\'nski, P. 1988, {\em J; Math.
Phys.\/}, {\bf 20}, 2576.
\par
Gruszczak, J. and Heller, M. 1993, {\em Intern. J. Theor. Phys.\/}
{\bf 32}, 625.
\par
Hartle, J. B. and Hawking, S.W. 1983, Wave Function of the Universe, \it
Phys. Rev. D\rm , \bf 28\rm , 2960-2975.
\par
Hasslacher, B. and Perry, M.J. 1981, Spin Networks Are Simplicial Quantum
Gravity, \it Phys. Lett.\rm , \bf 103B\rm , 21-24.
\par
Hawking, S. W. and Ellis, G. F. R. 1973, {\em The Large Scale
Structure of Space-Time\/}, Cambridge University Press: Cambridge.
\par
Hawking, S. and Penrose, R. 1966, {\em The Nature of Space-Time\/},
Princeton University Press: Princeton.
\par
Hawkins, E. 1996, {\em Hamiltonian Gravity and Noncommutative Geometry},
preprint, gr-qc/9605068.
\par
Heisenberg, W. 1930, Die Selbstenergie des Elektrons, \it Zeitschrift f\"ur
Physik\rm , \bf 65\rm , 4-13.
\par
Heller, M. 1992, Einstein Algebras and General Relativity,
 {\em Int. J; Theor. Phys.\/}, {\bf 31}, 277-288.
\par
Heller, M. and Sasin, W. 1995a, Structured Spaces and Their Applications
to Relativistic Physics, {\em J. Math. Phys.\/}, {\bf 36}, 3644-3662.
\par
Heller, M. and Sasin, W. 1995b, Sheaves of Einstein Algebras, {\em Int.
J. Theor. Phys.\/}, {\bf 34}, 387-398.
\par
Heller, M. and Sasin, W. 1996a {\em J. Math. Phys.}, in press.
\par
Heller, M. and Sasin, W. 1996b, {\em Banach Center Publications,\/} in=
 press.
\par
Jammer, M. 1974, \it The Philosophy of Quantum Mechanics. The
 Interpretations of
Quantum Mechanics in Historical Perspective\rm, Wiley, New York.
\par
Johnson, R. A. 1977, {\em J. Math. Phys.\/} {\bf 18}, 898.
\par
Joshi, P. S. 1993, {\em Global Aspects in Gravitation and
Cosmology\/}, Clarendon Press: Oxford.
\par
Joos, E. 1996, Decoherence Trough Interaction with the Environment, in:
\it Decoherence and the Appearance of a Classical World in Quantum
 Theory\rm,
Springer, Berlin, pp. 35-136.
\par
Kac, M. 1959, \it Probability and Related Topics in the Physical
 Sciences\rm,
Interscience Pub., New York.
\par
Kalau, W. 1996, {\em Hamilton Formalism in Non-Commutative Geometry},=
 preprint.
\par
Kolb, E. W. and Turner, M. S. 1990, {\em The Early Universe\/},
Addison -- Wesley: Redwood City -- Menlo Park, etc.
\par
LaFave, N.J. 1993, A Step Toward Pregeometry I: Ponzano-Regge Spin Networks
and the Origin of Space-Time Structure in Four Dimensions, preprint
gr-qc/9310036.
\par
Leibniz, G., \it The Monadology\rm, in: \it Philosophical Writings\rm ,
ed. G.H.R. Parkinson 1973, Dent: London.
\par
Lerner, D. 1972, Techniques of Topology and Differential Geometry
in General Relativity, in: {\em Methods of Local and Global
Differential Geometry in General Relativity\/}, eds. D. Farnsworth,
J. Frank, J. Porter and A. Thompson, Springer-Verlag: Berlin --
Heidelberg -- New York.
\par
Lorente, M. 1994, Quantum Processes and the Foundation of Relational
 Theories
of Space and Time, in: \it Relativity in General\rm , Proceedings of the
Relativity Meeting '93, Salas, September 7-10, 1993, eds. J. Diaz Alonso and
M. Lorente Paramo, \'Editions Fronti\`eres: Gif sur Yvette, pp. 297-302.
\par
Lorente, M. 1995, A Realistic Interpretation of Lattice Gauge Theories, in:
\it Fundamental Problems in Quantum Physics\rm , eds. M. Ferrero and A.
van der Merwe, Kluwer Academic Publishers: Dordrecht, pp. 177-186.
\par
Mach, E., 1960, {\em The Science of Mechanics: A Critical and
Historical Account of Its Development\/}, Open Court: LaSalle.
\par
Madore, J. 1995, {\em An Introduction to Noncommutative Differential=
 Geometry
and Its Physical Applications}, Cambridge University Press: Cambridge.
\par
Masson, T. 1996, {\em G\'eometrie non commutative et
applications \`a la theorie des champs\/}, Th\`ese, Preprint,
The Erwin Schr\"odinger International Institute for Mathematical
Physics.
\par
Misner, C.W. 1969, Quantum Cosmology I, \it Phys. Rev.\rm , \bf 186\rm ,
1319-1327.
\par
Misner, C. W., Thorne, K. S. and Wheeler, J. A. 1973, \it Gravitation\rm ,
W.H. Freeman and Company: San Francisco.
\par
Misra, B. and Sudarshan, E.C.G. 1977, The Zeno Paradox in Quantum Mechanics,
\it Journal of Mathematical Physics\rm, \bf 18\rm, 756.
\par
Nelson, E. 1985, \it Quantum Fluctuations\rm, Princeton University Press,
Princeton.
\par
Omn\`es, R. 1994, \it The Interpretation of Quantum Mechanics\rm, Princeton
University Press, Princeton.
\par
Partridge, R. B. 1955, {\em 3K: The Cosmic Microwave Background
Radiation\/}, Cambridge University Press: Cambridge.
\par
Patton, C.M. and Wheeler, J.A. 1975, Is Physics Legislated by Cosmogony?
in: \it Quantum Gravity\rm, eds. C.J. Isham, R. Penrose and D.W. Sciama,
Clarendon Press: Oxford, pp. 538-605.
\par
Penrose, R. 1971, Angular Momentum: an Approach to Combinatorial Space-Time,
in: \it Quantum Theory and Beyond\rm, ed. T. Bastin, Cambridge University
Press: Cambridge, pp. 151-180.
\par
Ponzano, G. and Regge, T. 1968, Semiclassical Limit of Racah Coefficients,
in: \it  Spectroscopy and Group Theoretical
Methods in Physics\rm , ed. F. Bloch, North Holland: Amsterdam, pp. 1-58.
\par
Prigogine, I. 1980, \it From Being to Becoming\rm, Freeman, New York.
\par
Prigogine, I. 1995a, Why Irreversibility? The Formulation of Classical and
 Quantum
Mechanics for Nonintegrable Systems, \it Int. J. Quantum Chemistry\rm,
\bf 53\rm, 105.
\par
Prigogine, I. 1995b, Science of Chaos or Chaos in Science --- A Rearguard
 battle,
\it Physicalia Magazine\rm, \bf 17\rm, 213.
\par
Prigogine, I. and Elskens, Y. 1985, Irreversibility, Stochasticity and
 Non-Locality in Classical Dynamics, communicated by Y. Elskens.
\par
Prigogine, I. and Stengers, I. 1984, \it Order out of Chaos\rm, Heinemann,
 London.
\par
Rindler, W. 1977, {\em Essential Relativity\/}, Springer: New York
-- Heidelberg~--Berlin.
\par
Roos, M. 1994, {\em Introduction to Cosmology\/}, John Wiley:
Chicester -- New York, etc.
\par
Sakharov, A.D. 1967, Vacuum Quantum Fluctuations in Curved Space and the
Theory of Gravitation, \it Doklady Akad. Nauk S.S.S.R.\rm , \bf 177\rm ,
70-71. English translation in: \it Sov. Phys. Doklady\rm , \bf 12\rm,
1040-1041 (1968).
\par
Sasin, W. and Heller, M. 1995, {\em Acta Cosmol.\/} {\bf 21}, no 2, 235.
\par
Schmidt, B. G., A New Definition of Singular Points in General
Relativity, {\em Gen. Rel. Grav.\/}, {\bf 1}, 269-280.
\par
Snyder, H.S. 1947, Quantized Space-Time, \it Phys. Rev.\rm, \bf 71\rm,
38-41.
\par
Stephani, H. 1982, {\em General Relativity -- An Introduction to
the Theory of the Gravitational Field\/}, Cambridge University
Press: Cambridge -- London~-- New York, etc.
\par
Tipler, F., Clarke, C. J. S. and Ellis, G. F. R., 1980,
Singularities and Horizons, in: {\em General Relativity and
Gravitation\/}, vol. 2, ed. A. Held, Plenum: New York, pp. 97-206.
\par
Tonnelat, M-A. 1965, \it Les th\'eories unitaires de l'\'electromagn\'etisme
 et de la
gravitation\rm, Gauthier-Villars, Paris.
\par
Vilenkin, A. 1988, Quantum Cosmology and the Issue of the Initial State of
 the
Universe, \it Phys. Rev.\rm , \bf 37\rm, 888-897.
\par
von Weizs\"acker, C.F. 1986, Reconstruction of Quantum Mechanics, in:
\it Quantum Theory and the Structure of Space and Time\rm , eds. L. Castell
and C.F. von Weizs=84ecker, Hanser: Munich.
\par
Wheeler, J. A. 1962, \it Geometrodynamics\rm , Academic Press: New York.
\par
Wheeler, J.A. 1964, Geometrodynamics and the Issue of the Final State,
in: \it Relativity, Groups and Topology\rm , eds. C. DeWitt and B.S. Dewitt,
Gordon and Breach: New York, pp. 315-320.
\par
Wheeler, J.A. 1968, Superspace and the Nature of Quantum Geometrodynamics,
in: \it Battelle Rencontres: 1967 Lectures in Mathematics and Physics\rm ,
eds. C. De Witt and J.A. Wheeler, W.A. Benjamin: New York, pp. 242-307.
\par
Wheeler, J.A. 1977, Genesis and Observership, in: \it Foundational
Problems in the Special Sciences\rm , eds. R. Butts and J. Hintikka,
Reidel: Dordrecht, pp. 1-33.
\par
Wheeler, J.A. 1979, Frontiers of Time, in: \it Problems in the Foundations
of Physics\rm, Proceedings of the International School of Physics ``Enrico
Fermi" (Course 72), ed. N. Toraldo di Francia, North-Holland: Amsterdam,
pp. 395-497.
\par
Wheeler, J.A. 1980, Pregeometry: Motivations and Prospects, in: \it Quantum
Theory and Gravitation\rm , ed. A.R. Marlow, Academic Press: New York, pp.
1-11.
\par
Wheeler, J.A. 1990, Information, Physics, Quantum, the Search
for Links, in: \it Complexity, Entropy and the Physics of Information\rm,
 ed:
W.H. Zurek, Addison-Wesley: Reading. See also: J. A. Wheeler 1994, It from
Bit, in:  \it At Home in the Universe\rm , The American Institute of
 Physics:
Woodbury, pp. 295-311.
\par

\end{document}